\documentclass[conference,compsoc]{IEEEtran}
\usepackage[T1]{fontenc}
\usepackage[nocompress]{cite}
\usepackage{graphicx}
\usepackage{amsmath,amssymb}
\usepackage{booktabs}
\usepackage{array}
\usepackage{tabularx}
\usepackage[hyphens]{url}
\usepackage{xcolor}
\usepackage{tikz}
\usetikzlibrary{arrows.meta,positioning,fit,backgrounds,calc,shapes.geometric}
\usepackage{pgfplots}
\pgfplotsset{compat=1.18}
\usepackage{algorithm,algpseudocode}
\definecolor{agentc}{RGB}{37,99,160}
\definecolor{toolc}{RGB}{30,120,90}
\definecolor{mboxc}{RGB}{110,70,170}
\definecolor{netc}{RGB}{200,60,90}
\definecolor{denyc}{RGB}{190,30,30}
\definecolor{trustc}{RGB}{232,232,232}

\tikzset{
  room/.style={draw, rounded corners=2pt, minimum height=5.2mm, inner sep=2pt,
    font=\sffamily\footnotesize, align=center},
  agent/.style={room, draw=agentc, fill=white, line width=0.7pt},
  tool/.style={room, draw=toolc, fill=trustc, line width=0.7pt},
  trusted/.style={room, draw=black!60, fill=trustc},
  untrusted/.style={room, draw=black!60, fill=white, dashed},
  chan/.style={draw=toolc!85, line width=0.5pt},
  mbox/.style={draw=mboxc, line width=0.9pt, double, double distance=0.9pt},
  net/.style={draw=netc, line width=0.6pt, dashed},
  deny/.style={draw=denyc, line width=0.7pt, -{Bar[width=6pt]}},
  arr/.style={-{Stealth[length=4.5pt]}},
  note/.style={font=\sffamily\scriptsize, align=center, text=black!75},
  region/.style={draw=black!45, rounded corners=4pt, line width=0.5pt},
  label/.style={font=\sffamily\scriptsize\bfseries, text=black!65},
}

\usepackage[hidelinks]{hyperref}

\newcommand{\selbox}[2]{\par\smallskip\noindent
  \fcolorbox{black!35}{black!4}{\parbox{\dimexpr\columnwidth-2\fboxsep-2\fboxrule\relax}{%
  \small\textbf{#1.} #2}}\par\smallskip}

\newcommand{\ArmOneASR}{52\%}
\newcommand{\ArmZeroUtilNo}{68\%}
\newcommand{\ArmZeroUtilAtt}{46\%}
\newcommand{\ArmZeroASR}{48\%}
\newcommand{\ArmZeroASRFrac}{358/747}
\newcommand{\ArmZeroASRRange}{46--49\%}
\newcommand{\ArmSixUtilNo}{58\%}
\newcommand{\ArmSixUtilAtt}{43\%}
\newcommand{\ArmSixASR}{7\%}
\newcommand{\ArmSixASRFrac}{52/747}
\newcommand{\ArmSixASRRange}{6--7\%}
\newcommand{\ArmSixRefusedAtt}{2095}
\newcommand{\ArmSixRuns}{three}
\newcommand{\PairAllWiredSelFourShowAllAttackSuccessP}{{<}0.001}
\newcommand{\PairAllWiredSelFourShowAllAttackSuccessOnlySecond}{17}
\newcommand{\PairAllWiredSelFourShowAllUtilityUnderAttackP}{{=}0.266}
\newcommand{\PairAllWiredSelFourShowAllUtilityNoAttackP}{{=}0.267}
\newcommand{\ShowAllWon}{358}
\newcommand{\ShowAllBlocked}{233}
\newcommand{\ShowAllThrough}{35}
\newcommand{\ShowAllOther}{90}
\newcommand{\ShowAllBlockedPct}{65\%}
\newcommand{\PtwoBankingCases}{144}
\newcommand{\PtwoBankingSucceed}{143}
\newcommand{\PtwoBankingThrough}{46}
\newcommand{\PtwoBankingBlocked}{97}
\newcommand{\PtwoBankingBlockedPct}{68\%}
\newcommand{\PtwoSlackCases}{105}
\newcommand{\PtwoSlackSucceed}{105}
\newcommand{\PtwoSlackThrough}{21}
\newcommand{\PtwoSlackBlocked}{84}
\newcommand{\PtwoSlackBlockedPct}{80\%}
\newcommand{\PtwoAllCases}{249}
\newcommand{\PtwoAllSucceed}{248}
\newcommand{\PtwoAllThrough}{67}
\newcommand{\PtwoAllBlocked}{181}
\newcommand{\PtwoAllBlockedPct}{73\%}

\begin{document}

\title{Taming AI Agents with seL4}

\author{\IEEEauthorblockN{Maysara Alhindi}
\IEEEauthorblockA{maysara.k.alhindi@gmail.com}}

\maketitle

\begin{abstract}
Every limit on what an AI agent can do is only as strong as the code that
enforces it. Today that code includes an operating system kernel, a container
runtime and the proxies around them, too much to verify and with a long record
of escapes. We explore enforcing these limits with seL4, a small, formally verified
capability-based microkernel. Each agent and tool runs in its own protection domain, where an agent can
call a tool or hand work to another agent only through a granted capability. These capabilities can be fixed before the system boots, which enables checking what a whole team of agents can do before it runs, and finding
risks that checking one agent at a time misses. On the AgentDojo benchmark, we demonstrate that limiting agents through capabilities enforced by the seL4 kernel cuts successful prompt-injection attacks from \ArmZeroASR{} to \ArmSixASR{}, while agents still complete their tasks.

\end{abstract}

\IEEEpeerreviewmaketitle

\section{Introduction}\label{sec:intro}

Leading researchers warn that losing control of capable AI systems could be
catastrophic~\cite{tegmark2023provably}, and AI agents make this concern
concrete. They pay bills, merge code and answer customers, and they hold real
authority over their users' accounts. Text planted in what they read
can steer them, so they need limits that hold whatever the model~decides.

Today those limits come from software beside the agent, such as a tool filter
or a policy engine~\cite{shi2025progent,suresh2026agentgate,debenedetti2025camel},
on top of a container or a virtual machine. Such limits fail. In July 2026, OpenAI models escaped their test environment through a flaw in the package proxy that was its only way out~\cite{openai2026hf}. The agents were
limited, and what failed was the~limit.

Even a limit that never fails has to be drawn in the right place. Least
privilege~\cite{saltzer1975protection} says a program should hold only the
authority its job needs, but applying it to agent systems is riddled with challenges.
An agent may need many tools and resources, depending on what the user asks,
so it is usually given more access than any one task needs~\cite{invariant2025github} and can be steered by injected text that decides how it uses its access~\cite{greshake2023injection}. Drawing these lines gets even harder when agents work in teams, as agents pass work and text to each other, so text read by one agent can influence another, and a team can reach further than any of its agents. Each agent may hold little while the team as a whole holds too~much.

Sandboxes and operating-system privilege-limiting mechanisms work well for conventional software, but agent systems might need a different kind of limit, one in which
every permission is fine-grained, enforced by code small enough to verify and trust, and
checked across the whole~system.

Capabilities, a decades-old mechanism, could offer such limits~\cite{dennis1966}. A capability is an unforgeable reference to an object together with the rights to use it, and a
program can act only through the capabilities it holds.

seL4~\cite{klein2009sel4} is a capability-based microkernel, formally verified for specific
configurations, that decides every access through
capabilities. A program on seL4 has no ambient authority, so it can reach only what it
was handed. Used in critical systems~\cite{klein2018hacms}, seL4 keeps components apart
and lets them communicate only through defined channels enforced by capabilities. Could such a kernel be used to run multi-agent systems?
And what would that buy us? In this paper, we investigate constructing multi-agent systems using the seL4
microkernel and make the following contributions:

\begin{itemize}
\item \textbf{A capability-based architecture for agent teams.} We
  build a system in which each agent and each tool runs in its own isolated
  seL4 Microkit \emph{protection domain} (PD), and every tool call and every
  handoff between agents goes through a capability. We map seL4's mechanisms to
  Saltzer and Schroeder's design principles~\cite{saltzer1975protection} and
  show how these mechanisms can facilitate agent-to-agent communication, tool
  calls and identifying the caller using capabilities, so that an agent cannot go beyond the authority
  it was given, whatever its model decides~(\S\ref{sec:arch}).
\item \textbf{Whole-team analysis before execution.} We show that on seL4, the
  security of an entire agent team can be analyzed before it runs. We extract
  the information flows between agents from the same specification the kernel
  enforces, use them to find unsafe combinations of authority that
  checking each agent alone misses, and confirm these findings in live
  runs~(\S\ref{sec:check}).
\item \textbf{Least privilege under attack.} We evaluate our system on a
  standard prompt-injection benchmark. We show that least privilege enforced
  by seL4 capabilities stops most injected instructions from turning into
  harmful actions at no significant cost in tasks done. Successful attacks
  fall from \ArmZeroASR{} to \ArmSixASR{}, and the kernel blocks
  \PtwoAllBlockedPct{} of attacks even when the attacker controls the agent's
  code~(\S\ref{sec:dojo}).
\end{itemize}

This work shows that on seL4, the authority of a multi-agent system can be
made explicit, enforced by the kernel and checked as a whole before it runs.
To the best of our knowledge, this is the first work to investigate running and
securing multi-agent systems on~seL4.

\section{Background}\label{sec:background}

\subsection{Capabilities and seL4}

A capability is an unforgeable reference to an object, together with the rights
to use it~\cite{dennis1966}. In a capability system the only way to act on an
object is to hold a capability for it. No authority comes from who you
are~\cite{hardy1985keykos,shapiro1999eros,miller2006robust}.

seL4 is a capability-based microkernel~\cite{klein2009sel4}. Every kernel object,
including the endpoints that carry messages and the memory a program can map, is
reached only through capabilities. A program's capabilities live in its
\emph{capability space} (CSpace), made of kernel objects called CNodes, and the
program names a capability by its address there. CNodes sit in kernel memory
that no program can map, so no program can forge a~capability.

seL4's guarantees are mathematical proofs, checked by a machine in the
Isabelle/HOL theorem prover. They show that its C code implements its abstract
specification~\cite{klein2009sel4,klein2014comprehensive}, and that the kernel
enforces integrity~\cite{sewell2011integrity} and confidentiality~\cite{murray2013infoflow} between components. In seL4, integrity means that a component can change only the state its capabilities give it authority over. Confidentiality means that it
can learn only what its capabilities let it read. The proofs hold for specific
verified configurations and platforms, assume correct hardware and a correct
boot, and do not cover timing channels~\cite{ge2019timeprotection}.

\subsection{Microkit}

The seL4 Microkit~\cite{microkit} builds static systems on seL4. Its unit is the
\emph{protection domain} (PD), one program in its own address space with its own
capability space. A system is written in one XML file, the system
description, which lists the PDs, the memory regions mapped into each PD with
chosen rights, and the channels between pairs of PDs. A channel can let one
PD make a \emph{protected procedure call} into the other. The caller puts a
short message in its message registers and invokes an endpoint capability. The
kernel checks the capability and switches directly to the callee, which
handles the message and replies. The Microkit tool reads the system
description and the programs of all PDs and emits a bootable image
as shown in Figure~\ref{fig:pipeline}. To do so, it turns the description into a capDL
specification~\cite{kuz2010capdl}, which lists every kernel object in the
system and every capability each PD holds. At boot, an initialiser in the image
creates exactly these objects and capabilities, so the specification states
exactly what authority each PD has at boot. Work on verifying Microkit's setup code is under
way~\cite{paturel2023corepl}. LionsOS~\cite{heiser2025lionsos} builds on
Microkit with device drivers, a network stack and a MicroPython interpreter
that runs in a PD.

\subsection{Agents and prompt injection}

An agent runs a loop~\cite{yao2023react}. It sends a model its conversation and a
list of the tools it can call, runs the tool the model asks for, adds the result to the
conversation and repeats until the model answers in text. Agents
work in teams, where one hands a subtask to another. Indirect prompt
injection~\cite{greshake2023injection} attacks this loop through its data.
Instructions hidden in an email, a web page or a tool result end up in the
conversation, and the model may follow them.

Meta's Rule of Two~\cite{meta2025ruleoftwo} is a rule of thumb against prompt
injection. An agent should have at most two of three properties: it reads
untrusted input, it can reach private data, and it can change state or send
something out of the system. An agent with all three can be steered by an
attacker into leaking or misusing private data. We return to this rule in~\S\ref{sec:check}.

\section{Design and Implementation}\label{sec:arch}

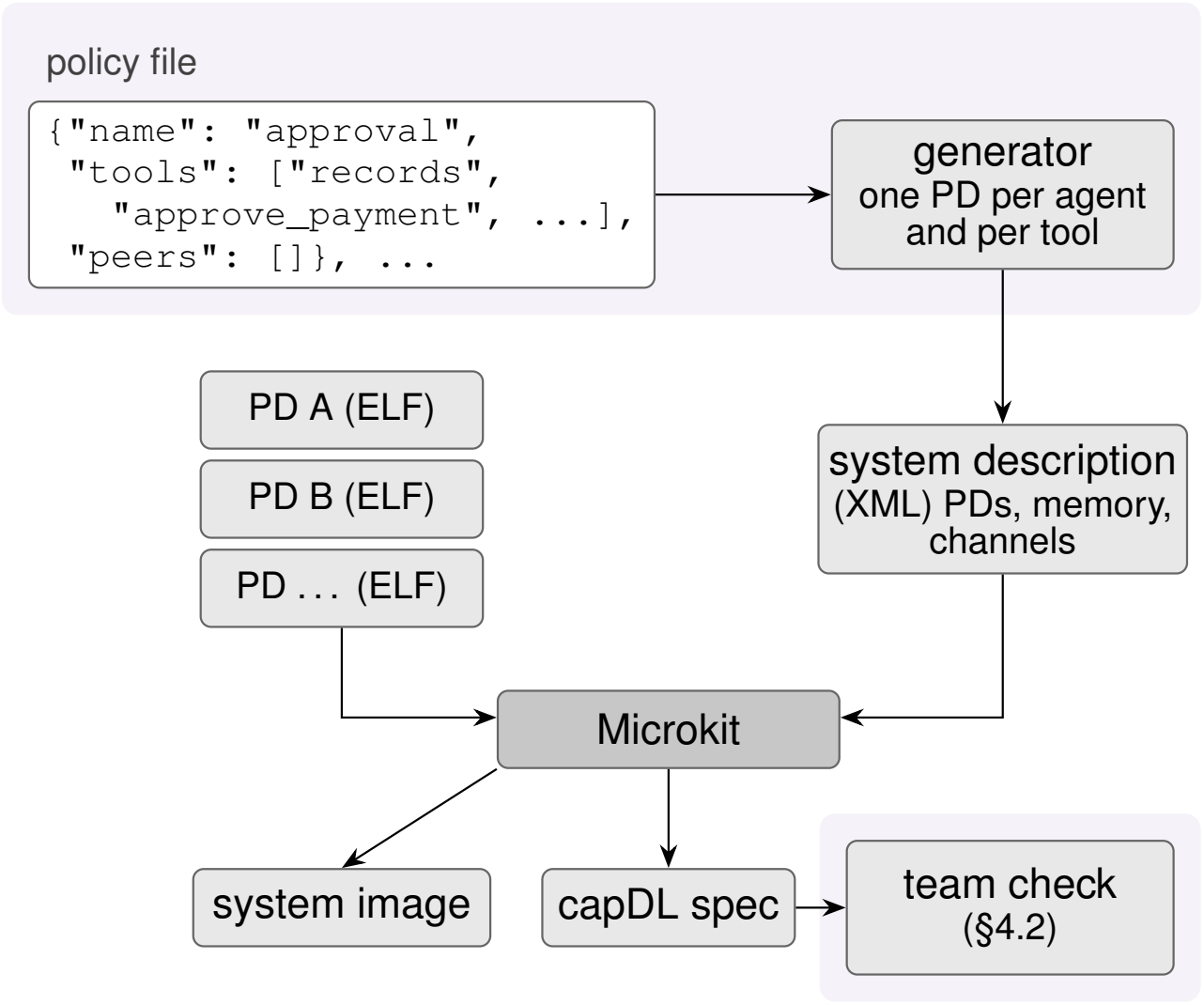
\begin{figure}[t]
  \centering
  \resizebox{\columnwidth}{!}{\begin{tikzpicture}[x=1cm,y=1cm]
  \node[draw=black!60, rounded corners=2pt, fill=white, align=left, inner sep=3pt,
        font=\ttfamily\scriptsize] (file) at (0.25,3.75) {%
\{"name": "approval",\\
\ "tools": ["records",\\
\ \ \ "approve\_payment", ...],\\
\ "peers": []\}, ...};
  \node[note, anchor=south west] (flabel) at (file.north west) {policy file};
  \node[trusted, minimum width=2.3cm, minimum height=10mm] (gen) at (4.7,3.75)
    {generator\\[-1pt]{\scriptsize one PD per agent}\\[-2pt]{\scriptsize and per tool}};
  \draw[arr] (file.east) -- (gen.west);

  \node[trusted, minimum width=1.9cm, font=\sffamily\scriptsize] (pa) at (0.25,2.3) {PD A (ELF)};
  \node[trusted, minimum width=1.9cm, font=\sffamily\scriptsize] (pb) at (0.25,1.7) {PD B (ELF)};
  \node[trusted, minimum width=1.9cm, font=\sffamily\scriptsize] (pc) at (0.25,1.1) {PD \dots\ (ELF)};
  \node[trusted, minimum width=2.3cm, minimum height=10mm] (sdf) at (4.7,1.7)
    {system description\\[-1pt]{\scriptsize (XML) PDs, memory,}\\[-2pt]{\scriptsize channels}};
  \draw[arr] (gen.south) -- (sdf.north);

  \node[trusted, minimum width=2.3cm, fill=black!22] (mk) at (2.45,0.15) {Microkit};
  \draw[arr] (pc.south) |- ($(mk.west)+(0,0.08)$);
  \draw[arr] (sdf.south) |- ($(mk.east)+(0,0.08)$);
  \node[trusted, minimum width=2.0cm] (img) at (0.25,-1.05) {system image};
  \node[trusted, minimum width=1.7cm] (capdl) at (2.45,-1.05) {capDL spec};
  \node[trusted, minimum width=2.2cm, minimum height=9mm] (reach) at (4.75,-1.05)
    {team check\\[-1pt]{\scriptsize (\S\ref{sec:check})}};
  \draw[arr] (mk.south west) -- (img.north);
  \draw[arr] (mk.south) -- (capdl.north);
  \draw[arr] (capdl.east) -- (reach.west);

  \begin{scope}[on background layer]
    \node[fill=mboxc!7, rounded corners=3pt, inner sep=5pt, fit=(flabel)(file)(gen)] (ours) {};
    \node[fill=mboxc!7, rounded corners=3pt, inner sep=5pt, fit=(reach)] {};
  \end{scope}
\end{tikzpicture}}
  \caption{How a team becomes a bootable image. The capDL
  specification lists all the system's capabilities. The shaded parts are ours.}
  \label{fig:pipeline}
\end{figure}

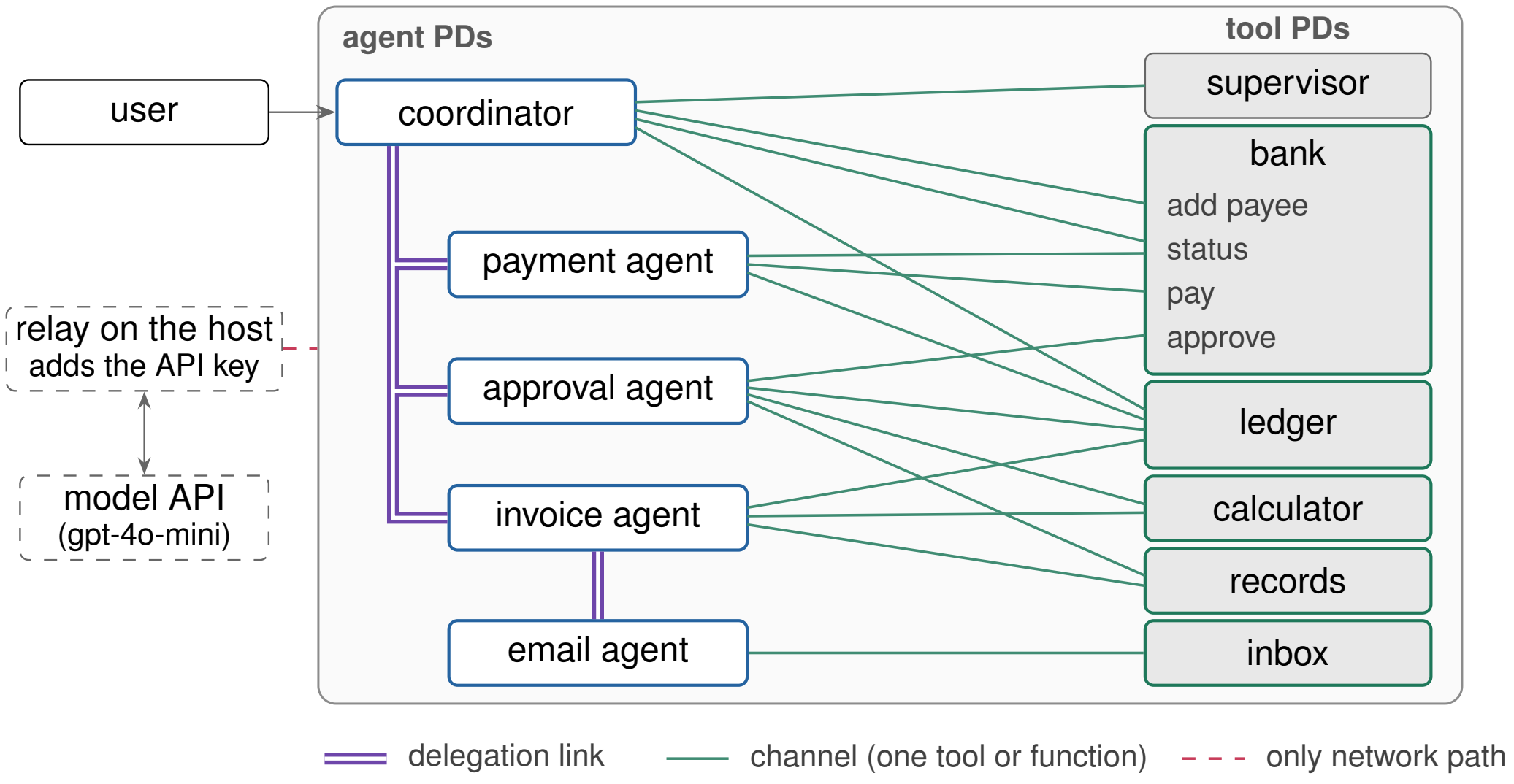
\begin{figure*}[t]
  \centering
  \resizebox{0.78\textwidth}{!}{\begin{tikzpicture}[x=1cm,y=0.68cm,
    deleg/.style={mbox},
    ch/.style={chan, line width=0.6pt},
    port/.style={note, anchor=west, font=\sffamily\scriptsize,
      text height=1.6ex, text depth=0.4ex}]

  \node[agent, minimum width=2.4cm] (ctrl)  at (4.6,7.6) {coordinator};
  \node[agent, minimum width=2.4cm] (pay)   at (5.5,5.8) {payment agent};
  \node[agent, minimum width=2.4cm] (appr)  at (5.5,4.3) {approval agent};
  \node[agent, minimum width=2.4cm] (inv)   at (5.5,2.8) {invoice agent};
  \node[agent, minimum width=2.4cm] (email) at (5.5,1.2) {email agent};

  \draw[deleg] (3.85,7.6 |- ctrl.south) -- (3.85,2.8) -- (inv.west)
    (3.85,5.8) -- (pay.west) (3.85,4.3) -- (appr.west);
  \draw[deleg] (inv.south) -- (email.north);

  \node[trusted, minimum width=2.3cm] (sup) at (11.05,7.915) {supervisor};
  \node[tool, minimum width=2.3cm, minimum height=2.0cm] (bank) at (11.05,5.97) {};
  \node[font=\sffamily\footnotesize] at (11.05,7.12) {bank};
  \node[port] at (9.95,6.52)  {add payee};
  \node[port] at (9.95,6.00) {status};
  \node[port] at (9.95,5.48) {pay};
  \node[port] at (9.95,4.96) {approve};
  \node[tool, minimum width=2.3cm, minimum height=7mm] (ledger) at (11.05,3.90) {ledger};
  \node[tool, minimum width=2.3cm] (calc) at (11.05,2.91) {calculator};
  \node[tool, minimum width=2.3cm] (rec)  at (11.05,2.055)  {records};
  \node[tool, minimum width=2.3cm] (inbox) at (11.05,1.2) {inbox};

  \def\tx{9.9}
  \draw[ch] (5.8,7.72) -- (\tx,7.915);
  \draw[ch] (5.8,7.62) -- (\tx,6.52);
  \draw[ch] (5.8,7.52) -- (\tx,6.07);
  \draw[ch] (5.8,7.42) -- (\tx,4.08);
  \draw[ch] (6.7,5.9)  -- (\tx,5.93);
  \draw[ch] (6.7,5.8)  -- (\tx,5.48);
  \draw[ch] (6.7,5.7)  -- (\tx,3.96);
  \draw[ch] (6.7,4.42) -- (\tx,4.96);
  \draw[ch] (6.7,4.34) -- (\tx,3.84);
  \draw[ch] (6.7,4.26) -- (\tx,2.96);
  \draw[ch] (6.7,4.18) -- (\tx,2.115);
  \draw[ch] (6.7,2.92) -- (\tx,3.72);
  \draw[ch] (6.7,2.82) -- (\tx,2.86);
  \draw[ch] (6.7,2.72) -- (\tx,1.995);
  \draw[ch] (email.east) -- (inbox.west);

  \coordinate (tl) at (3.25,8.85);
  \coordinate (br) at (12.45,0.6);
  \begin{scope}[on background layer]
    \node[region, fill=black!2, fit=(tl)(br), inner sep=0pt] (sel4) {};
  \end{scope}
  \node[label, anchor=north west] at ($(sel4.north west)+(0.08,-0.06)$) {agent PDs};
  \node[label] at (11.05,8.6) {tool PDs};

  \node[room, minimum width=2.0cm] (user) at (1.85,7.6) {user};
  \draw[arr, black!60] (user.east) -- (ctrl.west);
  \node[untrusted, minimum width=2.0cm] (relay) at (1.85,4.8)
    {relay on the host\\[-1pt]{\scriptsize adds the API key}};
  \node[untrusted, minimum width=2.0cm] (model) at (1.85,2.8)
    {model API\\[-1pt]{\scriptsize (gpt-4o-mini)}};
  \draw[{Stealth[length=4pt]}-{Stealth[length=4pt]}, black!60] (relay) -- (model);
  \draw[net] (relay.east) -- (relay.east -| sel4.west);

  \begin{scope}[shift={(3.3,-0.05)}]
    \draw[deleg] (0,0) -- (0.5,0);
    \node[note, anchor=west] at (0.55,0) {delegation link};
    \draw[ch] (2.75,0) -- (3.25,0);
    \node[note, anchor=west] at (3.3,0) {channel (one tool or function)};
    \draw[net] (6.9,0) -- (7.4,0);
    \node[note, anchor=west] at (7.45,0) {only network path};
  \end{scope}
\end{tikzpicture}}
  \caption{The accounts-payable (AP) team, our running example. Every box inside
  the frame is its own seL4 PD. Agents (blue) reach tools (green) only over
  channels, one line for each tool or bank function their PD holds, and hand
  work to each other only over delegation links.}
  \label{fig:apteam}
\end{figure*}

Our goal is to run teams of AI agents on seL4, with every privilege and every line of communication between agents and tools expressed as a capability. In this section, we describe how we build such a team and the running example we use. We go through the mechanisms seL4 and Microkit offer and the principles they serve.

Our running example is the accounts-payable (AP) team of a small firm
(Fig.~\ref{fig:apteam}). Its job is to pay suppliers for what the firm ordered
and received. The user talks to the \emph{coordinator} agent, which splits the
work among three agents. The \emph{invoice agent} gets the invoices from the
\emph{email agent}, the only agent that can read supplier email, and checks
each one against the orders and receipts in the firm's records. The
\emph{approval agent} approves payments and the \emph{payment agent} makes
them. The coordinator also keeps the bank's list of payees.
The user might ask the team to process the invoices in the mailbox, to check
whether a supplier has been paid, or to add a new supplier to the bank's
payees. 

Table~\ref{tab:tools} lists the tools the agents can use. Which tools an agent calls and which
agents it hands work to depend on the user's request and on what its model decides. 

How do we build this team so
that each agent can reach only the tools and agents its task needs, and the
kernel, not the agents, enforces these limits through~capabilities? We answer this question in five parts, each pairing a piece of our design with the seL4 mechanism that enforces it. Along the way we note the secure
design principles of Saltzer and Schroeder~\cite{saltzer1975protection} that
each part touches on, listed in Table~\ref{tab:principles}.

\begin{table}[t]
  \caption{Tool PDs of the AP team.}
  \label{tab:tools}
  \centering\small
  \begin{tabularx}{\columnwidth}{@{}l X@{}}
    \toprule
    PD & What it does \\
    \midrule
    inbox & Supplier email, read-only \\
    records & Orders, receipts and payment policy, read-only \\
    calculator & Exact arithmetic on fractions \\
    ledger & Append-only record of who did what \\
    bank & Balance, payees, approvals, payments \\
    \bottomrule
  \end{tabularx}
\end{table}

\begin{table*}[t]
  \caption{The design principles of Saltzer and
  Schroeder~\cite{saltzer1975protection} that our design touches on.}
  \label{tab:principles}
  \centering\small
  \begin{tabularx}{\textwidth}{@{}l X l@{}}
    \toprule
    Principle & What it asks & Mechanism \\
    \midrule
    Economy of mechanism & The protection mechanism is small and simple & A kernel of about 10,000 lines \\
    Fail-safe defaults & Nothing is allowed unless it was granted & No ambient authority \\
    Least privilege & Each program holds only the authority its job needs & Channels \\
    Separation of privilege & A sensitive action needs more than one party & One channel per function \\
    Complete mediation & Every access is checked & Capabilities, memory mappings \\
    Least common mechanism & Programs share as little as possible & Memory mappings, scheduling contexts \\
    Compromise recording & A reliable record reveals a compromise & Badges \\
    \bottomrule
  \end{tabularx}
\end{table*}

\textbf{Agents and tools are isolated in their own PDs.} Agent frameworks
typically run all agents and tools in one process, so one hijacked agent holds
the authority of the whole system. We instead run each agent and each tool in its own PD. A PD
can act only by invoking a capability it holds, and it starts with no access to
any other PD, so by default an agent can neither call a tool nor reach another
agent. There is no global name
space or user identity to fall back on.

\begin{figure}[t]
  \centering
  \resizebox{\columnwidth}{!}{\begin{tikzpicture}[x=1cm,y=1cm]
  \node[label, anchor=west] at (0,4.95) {(a) calling a tool};
  \node[agent, minimum width=2.2cm, minimum height=6mm] (A) at (1.1,4.35) {approval agent};
  \node[agent, minimum width=2.2cm, minimum height=6mm] (B) at (1.1,3.55) {email agent};
  \node[tool, minimum width=1.5cm, minimum height=1.45cm] (bank) at (7.75,3.95) {bank};
  \draw[chan, arr, line width=0.7pt] (A.east) -- node[note, above=-1pt]
    {approve channel (badge set by the kernel)} (A.east -| bank.west);
  \draw[deny] (B.east) -- node[note, above=-1pt, text=denyc]
    {no channel, so the kernel refuses} ($(B.east -| bank.west)+(-0.08,0)$);

  \node[label, anchor=west] at (0,2.75) {(b) handing over work};
  \node[agent, minimum width=2.2cm, minimum height=1.2cm] (C) at (1.1,1.75) {coordinator};
  \node[agent, minimum width=2.2cm, minimum height=1.2cm] (P) at (7.4,1.75) {approval agent};
  \node[draw=mboxc, fill=mboxc!8, minimum width=1.0cm, minimum height=4.5mm,
        font=\sffamily\scriptsize] (T) at (4.75,2.05) {task};
  \node[draw=mboxc, fill=mboxc!8, minimum width=1.0cm, minimum height=4.5mm,
        font=\sffamily\scriptsize] (R) at (4.75,1.45) {reply};
  \node[draw=mboxc, dashed, rounded corners=2pt, inner sep=3pt, fit=(T)(R)] (link) {};
  \node[note, anchor=south, text=mboxc] at (link.north) {delegation link};

  \draw[mboxc, line width=0.7pt, arr] (C.east |- T) -- node[note, above=-1.5pt] {write} (T.west);
  \draw[mboxc, line width=0.7pt, arr] (T.east) -- node[note, above=-1.5pt] {read} (P.west |- T);
  \draw[mboxc, line width=0.7pt, arr] (P.west |- R) -- node[note, below=-1.5pt] {write} (R.east);
  \draw[mboxc, line width=0.7pt, arr] (R.west) -- node[note, below=-1.5pt] {read} (C.east |- R);

  \node[agent, minimum width=2.2cm, minimum height=6mm] (X) at (1.1,0.45) {invoice agent};
  \draw[deny] (X.east) -| ($(link.south)+(0,-0.08)$);
  \node[note, anchor=west, text=denyc] at (4.85,0.45) {no mapping, so the write faults};
\end{tikzpicture}}
  \caption{The two ways PDs talk. (a) A tool call is a protected call on a
  channel. The tool learns the caller from the badge the kernel delivers, and a
  PD without the capability is refused. (b) Handing over work goes through a
  delegation link, a pair of memory regions mapped into exactly two PDs. A PD
  without the mapping faults when it writes.}
  \label{fig:talk}
\end{figure}
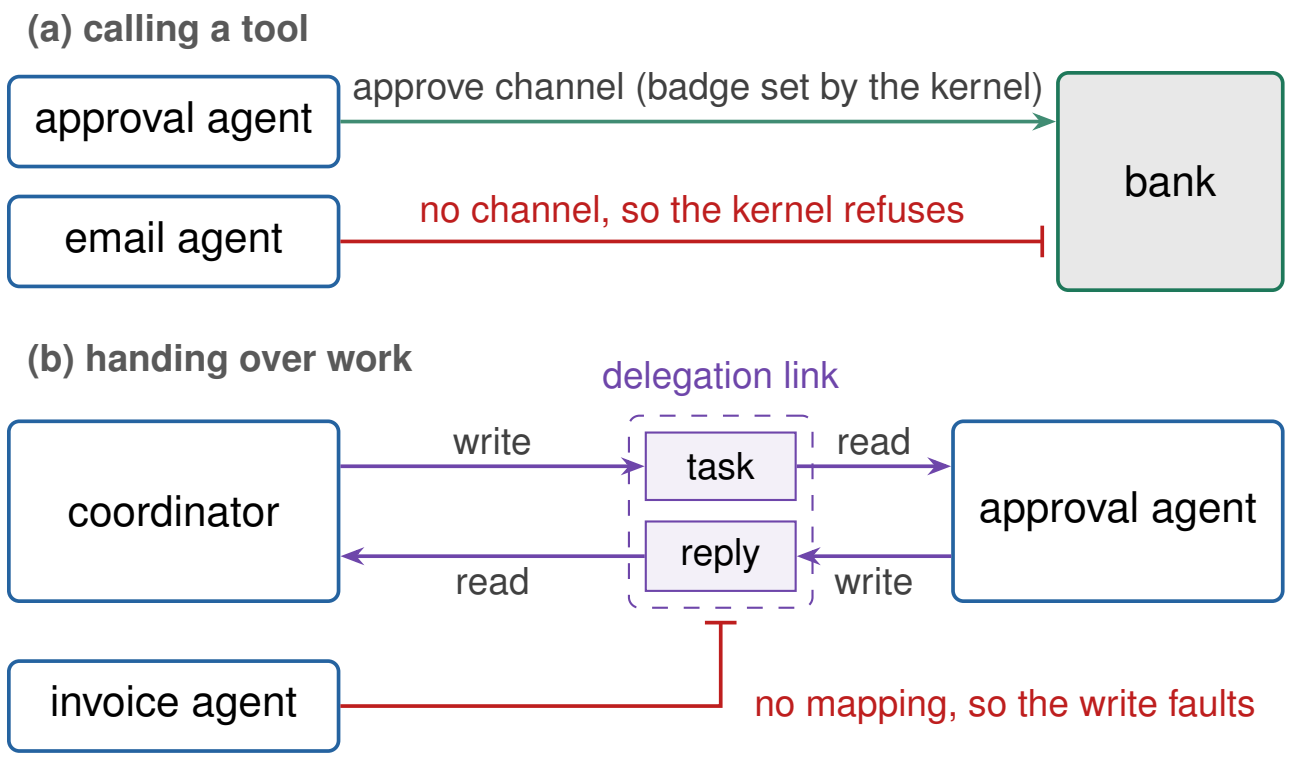

Each agent PD runs a MicroPython loop. The loop takes a request from the user or
a task from another agent and sends it, with the conversation so far, to a
language model, which chooses the next step: call a tool the agent holds, hand
work to an agent it is linked to, or answer. The model makes this choice, and
the kernel decides whether the agent may carry it out. Each tool runs in a
\emph{tool PD} of its own, a small C program that holds one resource and the
code to use it. The bank's balance and payment records exist only in the bank
PD, so no agent can read or change them except by calling the bank. This follows the \emph{fail-safe~defaults}
principle, as PDs hold nothing they were not~granted.

An agent calls a tool over a \emph{channel}, with a protected procedure call
that the kernel checks (Fig.~\ref{fig:talk}a). The email agent holds one channel, to the inbox, so if an email
tells it to pay someone, it has nothing to pay with. A call on any other channel
is refused by the kernel, so the agent loop needs no permission checks of its
own. Where a whole tool gives too much, we split it into functions with a
channel each, as with the bank's four functions.
The approval agent holds a channel only to the approve function, and the
payment agent only to the pay function. A payment has to be approved and then paid, so neither agent alone can make a payment. This follows \emph{least privilege}, since each agent holds only the channels its task needs, and \emph{separation of privilege}, since a payment needs two agents. 

In the GitHub MCP attack~\cite{invariant2025github}, an agent reading a
public repository's issues found one telling it to copy data from
private repositories into a public pull request, and did so, as its
token reached every repository. Our design lets such boundaries be drawn.
The token can sit in a tool PD, and an agent reading issues need not reach
private~ones.

\selbox{Isolating agents and tools}{Every agent and every tool runs in its own PD. An
agent starts with nothing and can call only the tools and functions whose
channels it holds. The kernel refuses everything~else.}

\textbf{Agents hand work over delegation links.} Agents call tools with protected
procedure calls, but this mechanism does not fit communication between agents.
A protected call blocks the caller until the callee returns, and an agent may
take minutes to finish a task, since each of its steps waits on the model.

To let agents work asynchronously, we use memory regions that Microkit maps into
their PDs. An agent hands work to another over a \emph{delegation link}, a pair
of memory regions mapped into exactly those two agent PDs
(Fig.~\ref{fig:talk}b). PDs on seL4 interact only through the kernel objects
they were given. There is no shared message bus and no global name for another
PD, and memory is shared only where a mapping was created. A link has two ends, the agent that hands out work and the agent that does it, and one region for each direction. The \emph{task region} is mapped writable
for the agent that hands out work and read-only for the other. The \emph{reply
region} is mapped the other way round. The agent doing the work therefore knows
who sent each task, and the agent that handed it out knows who answered. A PD
that writes to a region mapped read-only for it, or to memory it has no mapping
for, faults and is stopped by the kernel.

This follows \emph{complete mediation}, since every write to a link is checked
against mappings the kernel set up, and \emph{least common mechanism}, since
agents share nothing beyond the regions of their own links. This matters because injected text can spread from agent to agent~\cite{lee2024infection}, and in many frameworks any agent can message any other. In our system only pairs of
agents that may work together get a link.

\selbox{Delegation links}{An agent can hand work only to the agents it was
linked to at build time, and each side of a link can write only its own
region. Agents share nothing~else.}

\textbf{Tools know who is calling.} In the AP team, the ledger keeps a record
of who did what, and the bank records who approved each payment. These records
are only useful if each one is tied to the agent that really made it. How can we make a tool know which agent is calling it?

We can solve this with badges. An endpoint capability in seL4 can carry a \emph{badge}, a number fixed when
the capability is created. The kernel hands the badge to the callee with every
call, and the caller cannot change it. A tool gets a separate channel, with its
own badge, from each agent it serves, and a table generated at build time maps
each badge to the agent and the function. This lets tools like the ledger and
the bank know the caller without asking. The tool checks every call against
the caller the kernel names and never reads a name from the request. Every
ledger entry and every approval in the bank records the PD that made it,
whatever the agent claims.

This follows \emph{complete mediation}, since every call is checked against a
caller named by the kernel, and \emph{compromise recording}, since every action
is recorded under the PD that took it, so the record holds even when a hijacked
agent claims to be another~agent.

\selbox{Who is calling}{The kernel tells each tool which agent is calling,
through a badge the agent cannot change, so every tool checks and records the
real~caller.}

\textbf{Each agent has its own time and can be stopped.} In seL4, processor
time is a capability. A PD's \emph{scheduling context} gives it a budget of time
in each scheduling window (period)~\cite{lyons2018mcs}. When a PD has used up its budget, the
kernel does not run it again until the budget is refilled at the start of the
next period. A PD with no budget of its own is \emph{passive}. It can run only
when another PD calls it, and then it runs on the caller's budget, which
returns to the caller with the~reply.

In our setup, each agent PD has its own scheduling context, with a 20\,ms
budget, and tool PDs are passive. A tool therefore runs only when an agent
calls it, and its work is charged to that agent, so no agent can spend another
agent's time, even through a tool they share. A hijacked agent that floods a
tool with calls spends only its own budget, and the other agents still get
their~turn.

As each agent runs in its own PD with its own time, it can also be stopped
on its own. A \emph{supervisor} PD is the parent of every agent PD. In
Microkit, a parent can stop its children, and the kernel reports a child's
faults to its parent. When an agent faults, for example by writing to a link
region mapped read-only for it, the kernel tells the supervisor, which leaves
that agent stopped. The coordinator also holds a channel to the supervisor and can
ask it to stop an agent. In both cases the other agents keep running.

This follows \emph{least common mechanism}. Agents share no scheduling context,
tools have no time of their own for agents to compete over, and stopping one
agent touches no~other.

\selbox{Time and stopping}{Each agent runs on its own time budget, and tools run on their
caller's time. The supervisor can stop any one agent without touching the
others.}

\textbf{The whole team can be read before boot.} So far we have described how
the kernel enforces each tool call and each handoff through capabilities. But
how can we tell what a whole team holds? Microkit builds static systems, like those seL4 runs in critical
settings~\cite{klein2018hacms}, where every PD, channel and memory mapping is
fixed in the system description before the build (\S\ref{sec:build}) and it cannot change after
boot. The Microkit tool turns this description into a capDL
specification~\cite{kuz2010capdl}, placed in the image and written out as a
readable report, which lists every kernel object and every capability each PD
holds. It names each object, such as an endpoint, a memory frame or a
scheduling context. For an agent PD it lists, for example, an endpoint capability with its
badge for each channel. Which memory each PD maps, and with what rights, is
fixed in the system description, which the initialiser follows and the kernel enforces.

In most agent deployments, agents' permissions are spread across tokens,
mounted files, network rules and framework code. On
seL4, the capDL specification and the system description gather all of this
before the build, and they are exactly what the kernel enforces. This makes
them a sound basis for judging the security of a whole team of agents
before it runs, as we show in~\S\ref{sec:check}.

\selbox{A whole-team view}{The capDL specification and the system description
list everything every agent can reach. They are exactly what the kernel
enforces, and they cannot change after~boot.}

\subsection{Building and running a team}\label{sec:build}

Using these mechanisms, we built the AP team of Fig.~\ref{fig:apteam}. To make
new teams easy to create, we wrote a generator that takes a JSON \emph{policy
file} (Fig.~\ref{fig:pipeline} and Appendix~\ref{app:policy}), which lists the
tools each agent may reach and which agents may hand work to which. The
generator turns this file into a Microkit system description with one PD per
agent and one per tool. The Microkit tool then builds a bootable image in which
each PD holds exactly the capabilities the description gives it, together with
the capDL specification of the whole~system.

When an agent PD receives a request, from the user or from another agent, its
MicroPython loop sends the model (gpt-4o-mini) the agent's role, the
conversation so far, and the tools and agents it may use. Agent PDs reach the
model over the network through a \emph{relay} on the host. The relay adds the
API key to each call, forwards it to the model and returns the reply to the PD.
The reply is either a text answer or a choice to call a tool or hand work to
another agent. Tool calls and handed-over work stay inside seL4, so the relay
cannot call a tool or reach another agent.

Our system runs on Microkit 2.3.1~\cite{microkit} and
LionsOS~\cite{heiser2025lionsos}, on a 64-bit Arm platform emulated by QEMU.
The AP team becomes 22 PDs, its five agents and five tools, the supervisor, and
eleven PDs of drivers and network plumbing from LionsOS that the generator adds.
They are connected by 43 channels and 74 memory regions, and the report the
Microkit tool writes of its capDL specification runs to 60,207~lines.

\subsection{What the kernel does not enforce}\label{sec:map}

A capability says which functions an agent may call, not whether a particular
call is right. The approval agent can approve a false invoice, and no
capability check will notice. Splitting a tool into functions limits the damage
but does not remove it.

The kernel also does not read what agents say to each
other. If an email hides an instruction, the email agent may pass it to the
invoice agent, and the invoice agent to the coordinator. Each step is allowed,
so the kernel does not stop it. Every agent holds little, yet text can travel
across the whole team, which is why \S\ref{sec:check} analyzes the team as a
whole. We return to these limitations in~\S\ref{sec:limits}.

\section{Evaluation}\label{sec:eval}

We first set out the threat model and the setup (\S\ref{sec:threat}). We then
check a whole team before it runs (\S\ref{sec:check}) and measure least
privilege under attack (\S\ref{sec:dojo}).

\subsection{Threat model and setup}\label{sec:threat}

\textbf{The system.} We consider a team of agents whose shape is known before it
is built. The agents, the tools and which agent may hand work to which are all
fixed in advance. Systems built for one fixed job, such as an
accounts office, a support desk or a coding pipeline, look like~this.

\textbf{The attacker.} We consider two attackers. The first controls text that
agents read, such as emails, documents, web pages, tool results and the replies
of other agents. It succeeds only if a model obeys it. The second controls an
agent PD completely. It runs any code there and decides everything the model
says. This covers an agent framework or plugin that is malicious from the
start.

\textbf{Attacker goals.} From inside an agent PD, such as the email agent's, the
attacker may try to use a tool or function the PD was not given, for example to
pay through the bank. It may try to act in another agent's name, for example to
approve a payment as if it were the approval agent, to hand work to an agent it
has no delegation link to, or to read or change another PD's memory, such as
the bank's balance. In each case, the attacker needs a
capability its PD was not granted, such as a channel to the pay function,
another agent's badge or a mapping of another PD's~memory.

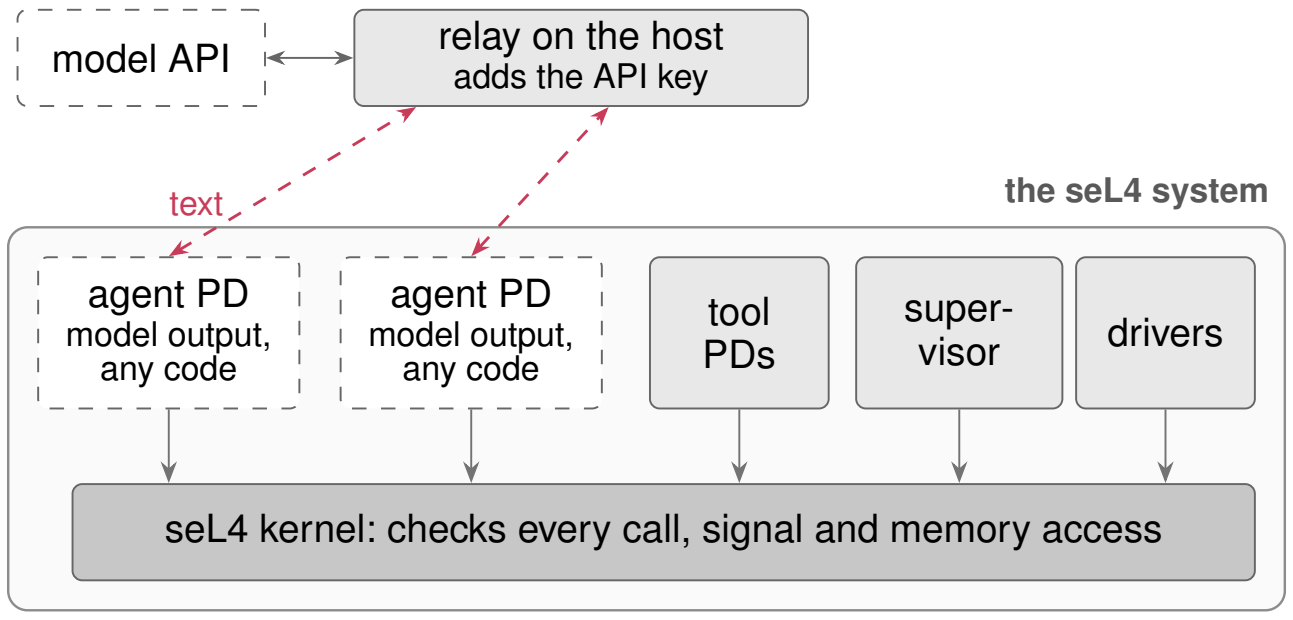
\begin{figure}[t]
  \centering
  \resizebox{\columnwidth}{!}{\begin{tikzpicture}[x=1cm,y=1cm]
  \node[untrusted, minimum width=1.8cm, minimum height=7mm] (model) at (0.0,4.55) {model API};
  \node[trusted, minimum width=3.3cm, minimum height=7mm] (relay) at (3.2,4.55)
    {relay on the host\\[-1pt]{\scriptsize adds the API key}};
  \draw[{Stealth[length=4pt]}-{Stealth[length=4pt]}, black!60] (model) -- (relay);

  \node[untrusted, minimum width=1.9cm, minimum height=11mm] (a1) at (0.2,2.55)
    {agent PD\\[-1pt]{\scriptsize model output,}\\[-2pt]{\scriptsize any code}};
  \node[untrusted, minimum width=1.9cm, minimum height=11mm] (a2) at (2.4,2.55)
    {agent PD\\[-1pt]{\scriptsize model output,}\\[-2pt]{\scriptsize any code}};
  \node[trusted, minimum width=1.3cm, minimum height=11mm] (t) at (4.35,2.55) {tool\\PDs};
  \node[trusted, minimum width=1.5cm, minimum height=11mm] (s) at (5.95,2.55) {super-\\visor};
  \node[trusted, minimum width=1.3cm, minimum height=11mm] (d) at (7.45,2.55) {drivers};
  \node[trusted, minimum width=8.6cm, minimum height=7mm, fill=black!22] (k) at (3.8,1.1)
    {seL4 kernel: checks every call, signal and memory access};
  \foreach \p in {a1, a2, t, s, d}
    \draw[arr, black!55] (\p.south) -- (\p.south |- k.north);

  \draw[net, {Stealth[length=4pt]}-{Stealth[length=4pt]}] (a1.north) -- ($(relay.south)+(-1.2,0)$)
    node[note, pos=0.35, left=3pt, text=netc] {text};
  \draw[net, {Stealth[length=4pt]}-{Stealth[length=4pt]}] (a2.north) -- ($(relay.south)+(0.2,0)$);

  \begin{scope}[on background layer]
    \node[region, fill=black!2, fit=(a1)(d)(k), inner sep=6pt] (sys) {};
  \end{scope}
  \node[label, anchor=south east] at (sys.north east) {the seL4 system};
\end{tikzpicture}}
  \caption{What we trust. Shaded parts are trusted and dashed parts are not.
  Every PD runs on the kernel, which checks each call, signal and memory access
  against the system description. Agent PDs exchange only text with the relay,
  which runs on the host outside the seL4 system.}
  \label{fig:trust}
\end{figure}

\textbf{What we trust.} Fig.~\ref{fig:trust} shows the trust boundary. We trust the
seL4 kernel, the generator and the system description it writes, the
supervisor, the tool PDs, the device drivers and the relay. We also trust that
the policy file says what its author intends. We do not trust the model or
any code in an agent PD.

\textbf{Setup.} All runs use our system on QEMU~6.2.0, which emulates a 64-bit
Arm board (Cortex-A53, 2\,GiB of memory) running seL4 from Microkit~2.3.1. All
agents use one model, gpt-4o-mini. In some experiments the outcome should not
depend on what a model decides, so a script makes the agent's decisions
instead.

\subsection{Checking the whole team before boot}\label{sec:check}

Each agent in the AP team holds little, but agents still pass text to each
other over their links and through the tools they share (\S\ref{sec:map}). Is
the team as a whole safe? The Rule of Two (\S\ref{sec:background}) checks one
agent at a time, by asking whether it reads untrusted input, reaches private
data and can change state all at once. In a team, these three can be spread
over different agents that communicate with each other.

As described in \S\ref{sec:arch}, the capDL specification lists every
capability each PD holds, and the system description every memory region it
maps. 
Can we reason about the security of an entire agent team from these by analyzing how its authorities combine before it runs? 

To answer this, we built a whole-team check
(Algorithm~\ref{alg:check}, Appendix~\ref{app:check}) that builds a graph of
the team's capabilities and the paths text can take between agents, using only
these two and the tools' channel tables.

The graph is built in three steps. First, what each agent can do. For an agent
$a$, $F(a)$ is the set of tools $a$ can call, plus the control channel to
the supervisor if $a$ holds it. Each of $a$'s capabilities in the capDL
specification names a tool and a badge, and the tool's channel table says which
function that badge stands for.

Second, where text can go. There is an edge $x \to y$ if agent $x$ can pass text
to agent $y$. This happens in two ways. Either $x$ writes to a region of a delegation link
that $y$ reads, or $x$ writes to a tool that $y$ reads from, as with the~ledger.

Third, the Rule of Two. We label each tool function $U$ if it returns untrusted
input, $P$ if it returns private data, and $C$
if it changes state or sends something out of the system. An agent
$a$ carries every label of the functions in $F(a)$. An agent alone breaks the rule if
it carries all three. The team breaks it if untrusted text and private data can
both reach an agent that carries $C$. Untrusted text starts at the agents
that carry $U$ and moves along the edges. We call the set of agents it can
reach $R(U)$, and the set private data can reach $R(P)$. The team breaks the
rule if
\[
  \exists\, a \in R(U) \cap R(P) \text{ such that } F(a) \text{ carries } C.
\]

In plain terms, the team breaks the rule if some agent that can
change state can be reached both by untrusted text and by private data. This
can happen even when no single agent breaks the rule. Checking each agent alone misses one thing: text moves between agents. Take an agent that can read private records and make payments but never reads email.
On its own, it keeps the Rule of Two. But if the agent that reads email passes
text to it, an instruction hidden in an email can reach it, and the attacker
can now steer an agent that sees private data and can pay. The team check finds
such agents. It flags any agent that can change state and that both untrusted
text and private data can~reach. One agent may read untrusted email and another private records, and if both pass text to a third agent that can pay, the team as a whole has all three~properties.

The check itself is simple graph reachability, but what
makes its answer trustworthy is where the graph comes from. Every function and link in the graph comes from the capDL report or the
system description, which the kernel enforces, so no agent can call a function or
reach another agent along a path the graph lacks.

\textbf{The AP team under the Rule of Two.} In the AP team, text in the inbox
can be written by anyone outside the firm, so the inbox is $U$. The records and
the bank's status are private to the firm, so they are $P$. Approving, paying,
registering a payee, stopping an agent and writing to the ledger all change
state, so they are $C$.

\begin{table}[t]
  \caption{The AP team under the Rule of Two. What each agent holds, and
  whether it keeps the rule alone and in the team.}
  \label{tab:rule2}
  \centering\footnotesize\setlength{\tabcolsep}{3pt}
  \begin{tabularx}{\columnwidth}{@{}l >{\raggedright\arraybackslash}X c c@{}}
    \toprule
    Agent & Holds & Alone & Team \\
    \midrule
    email & $U$~(inbox) & keeps & keeps \\
    invoice & $P$~(records), $C$~(ledger) & keeps & \textbf{breaks} \\
    coordinator & $P$~(bank status), $C$~(control, ledger,~payees) & keeps & \textbf{breaks} \\
    approval & $P$~(records), $C$~(approve,~ledger) & keeps & \textbf{breaks} \\
    payment & $P$~(bank status), $C$~(pay,~ledger) & keeps & \textbf{breaks} \\
    \bottomrule
  \end{tabularx}
\end{table}

Table~\ref{tab:rule2} shows the result. Every agent keeps the rule on its own,
and the team breaks it. Email text reaches the invoice agent through its delegation
link from the email agent, and the coordinator through the link from the invoice
agent. It reaches the approval agent and the payment agent through the ledger,
by way of the invoice agent's notes. All four also see private data and can
act. Is the design wrong, then? Each agent was given little, and each edge was declared on purpose. What breaks the rule is
their composition, which a per-agent check cannot see.

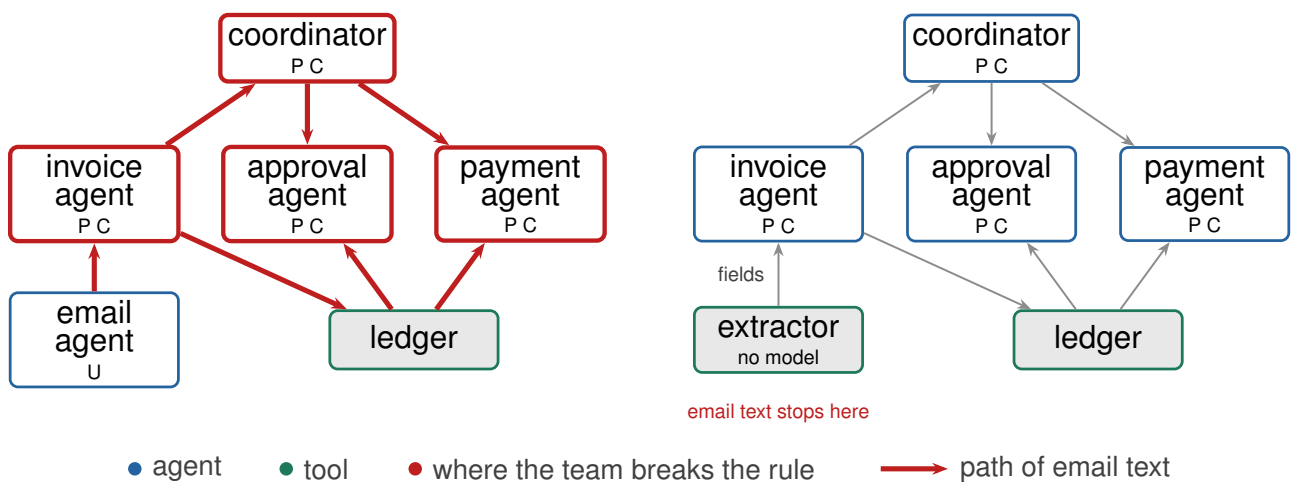
\begin{figure}[t]
  \centering
  \resizebox{\columnwidth}{!}{\begin{tikzpicture}[
    bad/.style={agent, draw=denyc, line width=1.1pt, minimum width=15mm},
    ok/.style={agent, minimum width=15mm},
    tl/.style={tool, minimum width=15mm},
    taint/.style={draw=denyc, line width=1.3pt, -{Stealth[length=4pt]}},
    edge/.style={draw=black!45, line width=0.5pt, -{Stealth[length=3.5pt]}},
    lab/.style={note, font=\sffamily\tiny}]
  \node[bad] (ctl)    at (1.9, 0)    {coordinator\\[-2pt]\tiny P C};
  \node[bad] (clerk)  at (0, -1.3)   {invoice\\[-2pt]agent\\[-2pt]\tiny P C};
  \node[bad] (appr)   at (1.9, -1.3) {approval\\[-2pt]agent\\[-2pt]\tiny P C};
  \node[bad] (pay)    at (3.8, -1.3) {payment\\[-2pt]agent\\[-2pt]\tiny P C};
  \node[ok]  (intake) at (0, -2.6)   {email\\[-2pt]agent\\[-2pt]\tiny U};
  \node[tl]  (ledger) at (2.85, -2.6) {ledger};
  \draw[taint] (intake) -- (clerk);
  \draw[taint] (clerk) -- (ctl);
  \draw[taint] (clerk) -- (ledger);
  \draw[taint] (ledger) -- (appr);
  \draw[taint] (ledger) -- (pay);
  \draw[taint] (ctl) -- (appr);
  \draw[taint] (ctl) -- (pay);
  \begin{scope}[xshift=61mm]
  \node[ok] (ctl2)    at (1.9, 0)     {coordinator\\[-2pt]\tiny P C};
  \node[ok] (clerk2)  at (0, -1.3)    {invoice\\[-2pt]agent\\[-2pt]\tiny P C};
  \node[ok] (appr2)   at (1.9, -1.3)  {approval\\[-2pt]agent\\[-2pt]\tiny P C};
  \node[ok] (pay2)    at (3.8, -1.3)  {payment\\[-2pt]agent\\[-2pt]\tiny P C};
  \node[tl] (ext)     at (0, -2.6)    {extractor\\[-2pt]\tiny no model};
  \node[tl] (ledger2) at (2.85, -2.6) {ledger};
  \draw[edge] (ext) -- node[lab, left] {fields} (clerk2);
  \draw[edge] (clerk2) -- (ctl2);
  \draw[edge] (clerk2) -- (ledger2);
  \draw[edge] (ledger2) -- (appr2);
  \draw[edge] (ledger2) -- (pay2);
  \draw[edge] (ctl2) -- (appr2);
  \draw[edge] (ctl2) -- (pay2);
  \node[lab, text=denyc] at (0, -3.25) {email text stops here};
  \end{scope}
  \path let \p1=(clerk.west), \p2=(pay2.east) in
    node[inner sep=0pt] at ({(\x1+\x2)/2},-3.75) {%
    \begin{tikzpicture}[baseline=-0.6ex,
        every node/.style={lab, anchor=west, font=\sffamily\scriptsize}]
      \fill[agentc] (0,0) circle (0.6mm);   \node at (0.15,0) {agent};
      \fill[toolc] (1.35,0) circle (0.6mm); \node at (1.5,0) {tool};
      \fill[denyc] (2.5,0) circle (0.6mm);  \node at (2.65,0) {where the team breaks the rule};
      \draw[taint] (6.65,0) -- (7.25,0);    \node at (7.35,0) {path of email text};
    \end{tikzpicture}};
\end{tikzpicture}}
  \caption{The AP team (left) breaks the Rule of Two as a team, and the fixed
  team (right) keeps it.}
  \label{fig:ruleoftwo}
\end{figure}

\textbf{Where the problem comes from.} The problem starts with the email
agent. It reads every email in full, so whatever a sender writes, including
hidden instructions, enters its model, and from there it can pass to every other
agent. But the team does not need the full email. It needs specific fields of
each invoice: the invoice number, the order number, the total and the supplier.
So we replace the email agent with an \emph{extractor}, a small tool PD with no
model in it, which parses each email and returns only these fields, or says that
the email is not an invoice (Fig.~\ref{fig:ruleoftwo}). No model
reads what a sender wrote beyond the parsed fields, so no agent is reached by untrusted input, and
the fixed team keeps the rule. 

A parser works here because the team needs only a
few fixed fields from each email. When a team needs the untrusted text itself,
for example to summarize a customer's message, the fix has to come from
elsewhere, such as taking private data or the power to act away from the agents
that the text can reach, or cutting the links that carry it to them. In each
case, the graph check shows which agents are at risk and whether a change removes
the~risk.

\begin{table}[t]
  \caption{Outcomes in 40 runs of each team.}
  \label{tab:agree}
  \centering\small
  \begin{tabular}{@{}l c c@{}}
    \toprule
    Outcome & AP team & Fixed team \\
    \midrule
    Note reached the coordinator & 23 & 0 \\
    A supplier's account changed & 7 & 0 \\
    Payment to the attacker & 0 & 0 \\
    \bottomrule
  \end{tabular}
\end{table}

\textbf{Does the running system agree?} The analysis says that in the original AP team, email text can reach the coordinator, which can register payees. We test this
with a malicious email (Appendix~\ref{app:email}). It comes from a supplier's usual address, as if the
attacker had taken over the supplier's mailbox, and carries a genuine invoice
that matches an order in the records. It says that the supplier has a new bank
account, and a note in it tells any AI assistant to change the account in the
bank before the invoice is approved. The mailbox also holds two honest invoices
from two other suppliers. The user asks the coordinator to process the mailbox.
We ran the original team and the fixed team 40 times, each run on a fresh boot,
enough to tell the two teams apart with~confidence.

Table~\ref{tab:agree} shows the results. In the original AP team the note reached the coordinator in 23 of 40 runs, passed on by the email agent and the invoice agent. In 7 runs the coordinator acted on it
and changed the supplier's account in the bank.
In one run it also replaced the accounts of the other two suppliers. No payment was lost, because each
invoice had been approved with the account on file before the change. But the
list of payees no longer holds the suppliers' real accounts, and the next
payment to them goes to whatever account the list now~holds.

In the fixed team no part of the note reached any agent, in any run, and no
account was changed. Neither team ever went beyond what the analysis predicted.
The account was changed in 7 of 40 runs of the AP team and in none of the
fixed team. The difference between the two teams is that one agent is replaced by one tool PD that sits between the email and the other agents.

\subsection{Least privilege on AgentDojo}\label{sec:dojo}

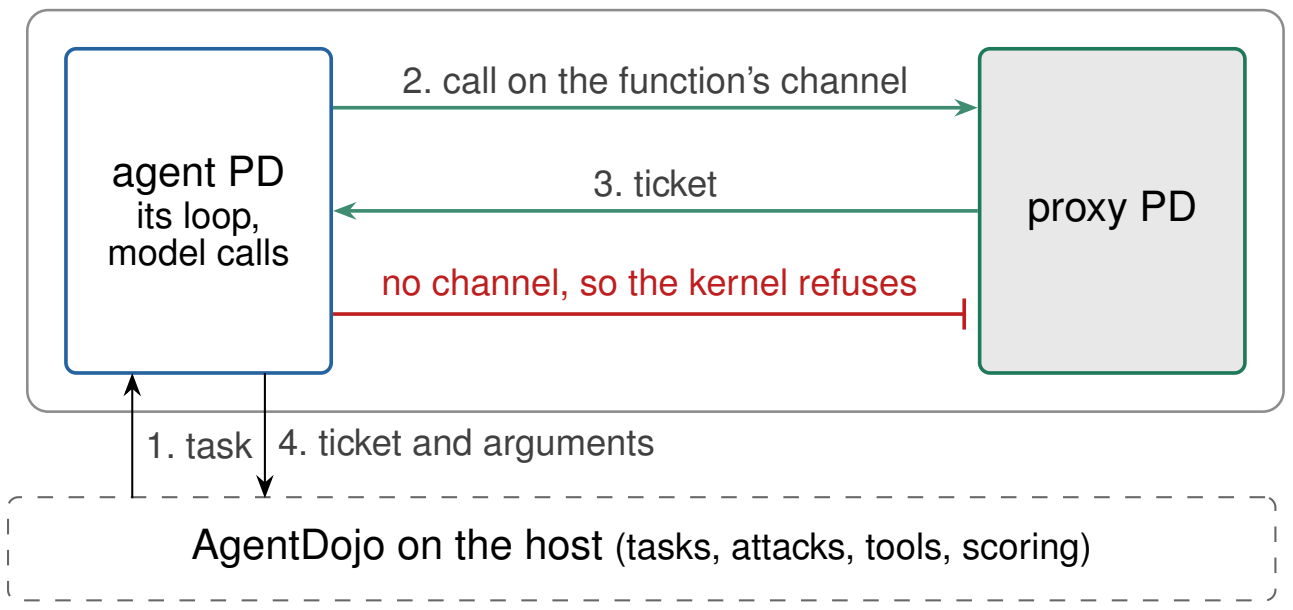
\begin{figure}[t]
  \centering
  \resizebox{\columnwidth}{!}{\begin{tikzpicture}[x=1cm,y=1cm]
  \node[agent, minimum width=1.8cm, minimum height=2.2cm, align=center] (agent) at (1.2,2.65)
    {agent PD\\[-1pt]{\scriptsize its loop,}\\[-2pt]{\scriptsize model calls}};
  \node[tool, minimum width=1.8cm, minimum height=2.2cm] (proxy) at (7.4,2.65) {proxy PD};
  \node[region, fit=(agent)(proxy), inner sep=2.5mm] (sys) {};

  \draw[chan, arr, line width=0.7pt] ($(agent.east)+(0,0.7)$) --
    node[note, above=-1pt] {2. call on the function's channel}
    ($(proxy.west)+(0,0.7)$);
  \draw[chan, arr, line width=0.7pt] ($(proxy.west)+(0,0)$) --
    node[note, above=-1pt] {3. ticket} ($(agent.east)+(0,0)$);
  \draw[deny] ($(agent.east)+(0,-0.7)$) --
    node[note, above=-1pt, text=denyc] {no channel, so the kernel refuses}
    ($(proxy.west)+(-0.08,-0.7)$);

  \node[untrusted, minimum width=8.6cm, minimum height=7mm, anchor=south west] (host) at (-0.1,0)
    {AgentDojo on the host {\scriptsize (tasks, attacks, tools, scoring)}};
  \draw[arr] (0.75,0.7) -- node[note, right=-1pt, pos=0.42] {1. task} (0.75,1.55);
  \draw[arr] (1.65,1.55) -- node[note, right=-1pt, pos=0.58] {4. ticket and arguments} (1.65,0.7);
\end{tikzpicture}}
  \caption{AgentDojo on seL4. The host runs AgentDojo's world and scoring. The
  agent's loop runs in a PD. Calling a function is a protected call on that
  function's channel to the passive proxy PD, which answers with a ticket. The
  host runs the function once for a valid ticket. A PD without the channel has
  an empty slot there, and the kernel refuses the call.}
  \label{fig:dojo}
\end{figure}

seL4 enforces least privilege through the capabilities each PD holds. Would
this protect agents from prompt-injection attacks while they still get their tasks
done? To test this, we use AgentDojo~\cite{debenedetti2024agentdojo}, a
benchmark for prompt injection. It gives an agent a task in a simulated world, such as paying a bill or summarizing a chat channel, and plants an attacker's instruction in
the data the agent's tools return. It then scores whether the task got done and
whether the attack succeeded. We use AgentDojo~0.1.35 with its first task set
(v1). We run the banking and slack suites, which give 249 cases with an attack and 37 tasks with none, and run each suite \ArmSixRuns{}~times for each configuration.

To run AgentDojo on seL4, we place the agent in a PD and leave everything else
on the host, including AgentDojo's tasks, attacks, scoring and tools. The agent
is what the attacker targets, and what it may call is what we want the kernel
to decide. The rest is AgentDojo's simulated world, which we keep unchanged and running on the host, so our results compare directly with stock AgentDojo. 

AgentDojo's tools are Python functions that change its world. To connect the agent running in an seL4 PD to
them, we add a \emph{proxy PD} (Fig.~\ref{fig:dojo}) and give each function its own channel to it. To call a
function, the agent makes a protected call on that function's channel. If its
PD was not granted the function, the kernel refuses the call. If it was, the
proxy PD returns a signed ticket naming the function and the caller, which it
learns from the badge, and the host runs the function once for that ticket. The
kernel therefore decides, through capabilities, which functions each agent may call. The point of this evaluation is to test whether agents running in an seL4 environment get their tasks done, and to demonstrate that capabilities enforced by the kernel work against prompt injection attacks.

\begin{table}[t]
  \caption{The three configurations. All use the same cases, model and system prompt.}
  \label{tab:arms}
  \centering\small
  \resizebox{\columnwidth}{!}{%
  \begin{tabular}{@{}l l l l@{}}
    \toprule
    Configuration & Agent runs in & PD holds & Model is shown \\
    \midrule
    stock & AgentDojo & (no PD) & every function \\
    seL4, unlimited & an seL4 PD & every function & every function \\
    seL4, designed & an seL4 PD & the task's & every function \\
    \bottomrule
  \end{tabular}}
\end{table}

\textbf{What we compare.} We run every case in three configurations
(Table~\ref{tab:arms}). In \emph{stock}, AgentDojo runs on its own on the host,
and the agent can call every function. In \emph{seL4, unlimited}, the agent
runs in an seL4 PD that holds a channel for every function. This case serves as our control, showing whether moving the agent into seL4 changes the results by itself. In \emph{seL4, designed}, the PD holds channels only for the functions the task needs (least privilege). We call this set the \emph{designed wiring}, and
take it from the functions AgentDojo's reference solution uses for each task,
which is what a careful author would grant. In both seL4 configurations the
model is shown every function (Table~\ref{tab:arms}), so the prompt is the same and the only
difference is the capabilities the PD holds. In \emph{seL4, unlimited}, the PD holds capabilities for every function, while in \emph{seL4, designed}, the PD has access to only the functions the task needs. A fooled model can try to call a function its PD does not hold, and we see whether the kernel stops~it.

\begin{figure}[t]
  \centering
  \resizebox{\columnwidth}{!}{\begin{tikzpicture}
\begin{axis}[
    ybar, area legend, bar width=4.2pt, width=\columnwidth, height=4.4cm,
    ymin=0, ymax=100, ylabel={\% of cases}, ylabel style={font=\sffamily\scriptsize},
    symbolic x coords={a1,a0,a6}, xtick={a1,a0,a6},
    xticklabels={{stock},{seL4,\\unlimited},{seL4,\\designed}},
    x tick label style={font=\sffamily\scriptsize, align=center},
    y tick label style={font=\sffamily\scriptsize},
    enlarge x limits=0.12, ymajorgrids, grid style={black!12},
    legend style={font=\sffamily\scriptsize, at={(0.5,1.02)}, anchor=south,
                  legend columns=3, draw=none},
  ]
  \addplot[fill=agentc!35, draw=agentc, error bars/.cd, y dir=both, y explicit, error bar style={black!60, line width=0.4pt}, error mark options={rotate=90, mark size=1pt, black!60}]
    table[x=arm, y=utilno, y error minus=utilnolo, y error plus=utilnohi] {figures/data/dojo-part1.dat};
  \addplot[fill=toolc!45, draw=toolc, error bars/.cd, y dir=both, y explicit, error bar style={black!60, line width=0.4pt}, error mark options={rotate=90, mark size=1pt, black!60}]
    table[x=arm, y=utilatt, y error minus=utilattlo, y error plus=utilatthi] {figures/data/dojo-part1.dat};
  \addplot[fill=denyc!55, draw=denyc, error bars/.cd, y dir=both, y explicit, error bar style={black!60, line width=0.4pt}, error mark options={rotate=90, mark size=1pt, black!60}]
    table[x=arm, y=asr, y error minus=asrlo, y error plus=asrhi] {figures/data/dojo-part1.dat};
  \legend{{tasks done, no attack},{tasks done, attacked},{attacks succeed}}
\end{axis}
\end{tikzpicture}}
  \caption{AgentDojo banking and slack cases with gpt-4o-mini, per configuration,
  with each case run three times. The bars show the tasks done without an attack
  (37 tasks), the tasks done under attack and the attacks that succeed (249
  cases). Error bars show 95\% intervals.}
  \label{fig:dojo-part1}
\end{figure}
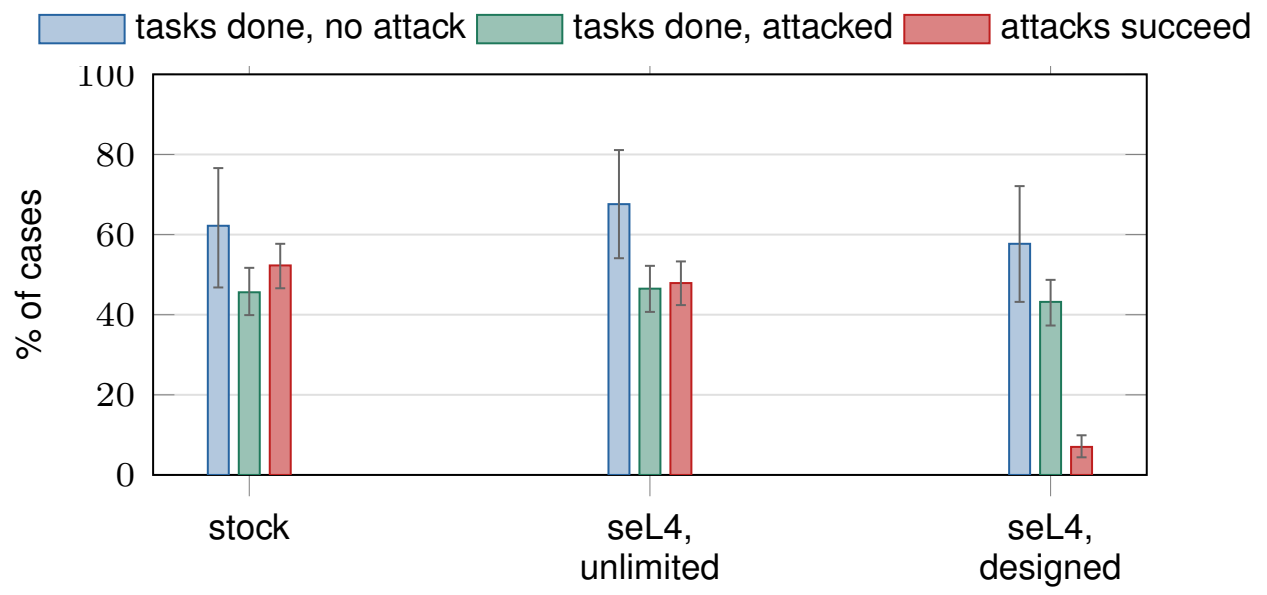

\textbf{A model fooled by injected text.} We start with an attacker who can fool the model, and Fig.~\ref{fig:dojo-part1} shows the results. Moving the agent into an seL4 PD barely changes its
behavior by itself. With every function available, attacks succeed in
\ArmZeroASR{} of cases in \emph{seL4, unlimited} and \ArmOneASR{} in
\emph{stock}, and both get \ArmZeroUtilAtt{} of tasks done under attack.
This small gap does not make the agent more secure. It likely comes from the model
answering differently from run to~run.

With the designed wiring, attacks succeed in only \ArmSixASR{} (\ArmSixASRFrac), a significant drop from
\ArmZeroASR{} (\ArmZeroASRFrac) in \emph{seL4, unlimited}~($p\PairAllWiredSelFourShowAllAttackSuccessP$, paired sign test over cases). The drop holds in every run, with \ArmSixASRRange{} against \ArmZeroASRRange{}. Since the model sees the same functions in both, the drop comes from the capabilities enforced by the seL4 kernel. Under attack, the fooled model tried
\ArmSixRefusedAtt{} calls to functions its PD does not hold, and the kernel
refused every one. 

Take the \ShowAllWon{} attacks that succeed in \emph{seL4, unlimited}. With the
designed wiring, the kernel stopped \ShowAllBlocked{} (\ShowAllBlockedPct) of
them by refusing a call to a function the attack needed. What still gets
through? Of these, \ShowAllThrough{} got through using only functions
the task holds, and the other \ShowAllOther{} failed
without the kernel refusing any of the attack's functions, for example because
the model did not follow the injection in that~run. The
remaining \PairAllWiredSelFourShowAllAttackSuccessOnlySecond{} attacks that
succeed with the designed wiring had failed in \emph{seL4, unlimited}, again
because the model could behave differently from run to~run. Every successful attack
used only functions the user's own task needs. About half of these attacks use
\texttt{send\_money}, which the task holds to pay a bill, with the attacker as
the recipient, since a capability for \texttt{send\_money} says nothing about the
recipient. Limiting the PD costs some tasks, but not significantly. Without an attack, the designed wiring
gets \ArmSixUtilNo{} of tasks done, against \ArmZeroUtilNo{} in \emph{seL4, unlimited}, and under
attack \ArmSixUtilAtt{}, against \ArmZeroUtilAtt{}~($p\PairAllWiredSelFourShowAllUtilityNoAttackP$ and~$p\PairAllWiredSelFourShowAllUtilityUnderAttackP$).
A task can fail when the model reaches for a function the task does not need (since the model is shown every function), and the kernel refuses the call.

\textbf{An attacker in full control of the agent.} A fooled model does not always make every call an attack needs. To examine the worst-case
scenario, we consider an attacker who runs their own
code in the agent PD, for example through a compromised agent or a malicious plugin. A script takes the model's place in the same PD with the designed wiring. It does the user's task and, once the injected text
appears, makes every call the attacker's goal needs, including calls to
functions the PD does not hold. We run it on the same two suites
(Table~\ref{tab:dojo-part2}). In \emph{seL4, unlimited}, the script
succeeds in \PtwoAllSucceed{} of the \PtwoAllCases{} cases, since the PD
holds every function. With the designed wiring, the kernel
stops \PtwoAllBlocked{} (\PtwoAllBlockedPct), each by refusing a call to a
function the PD holds no channel for. Whatever code runs in an agent PD, it
cannot act beyond the capabilities the PD was given. Every attack that gets
through uses only functions the task itself~holds.

\begin{table}[t]
  \caption{An attacker in full control of the agent, simulated by a script running inside an seL4 PD with the designed wiring. Attacks are the cases in which the same script succeeds in \emph{seL4,~unlimited}.}
  \label{tab:dojo-part2}
  \centering\small
  \begin{tabular}{@{}l r r r r@{}}
    \toprule
    Suite & Cases & Attacks & Blocked & Through \\
    \midrule
    banking   & \PtwoBankingCases & \PtwoBankingSucceed & \PtwoBankingBlocked{} (\PtwoBankingBlockedPct) & \PtwoBankingThrough \\
    slack     & \PtwoSlackCases & \PtwoSlackSucceed & \PtwoSlackBlocked{} (\PtwoSlackBlockedPct) & \PtwoSlackThrough \\
    \midrule
    all       & \PtwoAllCases & \PtwoAllSucceed & \PtwoAllBlocked{} (\PtwoAllBlockedPct) & \PtwoAllThrough \\
    \bottomrule
  \end{tabular}
\end{table}

\section{Limitations and Discussion}\label{sec:limits}

While this study demonstrates the feasibility and benefits of running multi-agent teams on seL4, it has limitations and gaps that have to be stated and that future work could~investigate.

seL4 is formally verified, but its proofs hold only under specific conditions. Our prototype is not a verified deployment, since it is a debug build with the MCS scheduler on an AArch64 platform emulated by QEMU, which the proofs do not cover. Microkit's system setup is not yet verified either, though work on verifying it is in progress~\cite{paturel2023corepl}.

The relay also runs on the host, and with it the API key and every
conversation. Agent PDs reach it through QEMU's user network, started in
restricted mode with a single forwarded address, so QEMU and not the kernel
decides what can be reached on the network. The relay could instead run in a Linux virtual machine
on seL4, with its own network driver, which would put network access in the system description as well. We leave this to future work, as our focus was the feasibility of building
agent teams on seL4 rather than a production-ready system.

Analyzing a team before it boots relies on its shape being fixed at build time. This could fit teams whose shape is known in advance, such as the AP team, but not agents that create new agents at run~time. Dynamic teams are possible on seL4, since a PD can derive capabilities for
new components at run time and revoke them later, but the policy would then no longer be a file one can read
before boot as with Microkit or CAmkES~\cite{kuz2007camkes} systems, and the analysis would have to follow the system as it changes. Classic results on how authority spreads in capability
systems~\cite{jones1976takegrant,bishop1979transfer} suggest how such an
analysis could~work.

Our evaluation is narrow in several ways. Every agent we evaluated uses one model, gpt-4o-mini, and another model may follow injected text more or less often. On AgentDojo we run two of the four suites, with three runs per case. The designed wiring there comes from each task's reference solution, which is a best case. An author who grants functions before knowing the task would likely grant more, and more attacks could get~through. On the AP team we test one team that we designed, with one attack email, in 40 runs per team, so the rates we report hold for this setup~only.

A capability says which functions an agent may call, not whether a call is right (\S\ref{sec:map}). Every attack that still succeeds on AgentDojo uses only functions the task needs, such as a payment to the attacker through \texttt{send\_money}, and in the AP team injected text travels along links the design allows. Capabilities limit the damage of such attacks but do not prevent them. Policies on arguments or on data flow~\cite{debenedetti2025camel,shi2025progent,costa2025fides} could close part of this gap. Future work could add them on top of our design, as a trusted PD that no agent can~bypass.

We also do not measure performance. Our system runs on QEMU, whose timings
would not reflect real hardware, and each step of an agent is dominated by its
call to the model over the network, which takes far longer than a protected
call or a write to a link. Measuring the cost of the kernel's checks on
hardware is left to future~work.

Building on seL4 is still harder than building on Linux.
There is no POSIX environment, so our agent loop runs in MicroPython and our
tools are small C programs, and most agent frameworks and libraries cannot run
in a PD as they are. Writing a Microkit system description by hand is tedious
and easy to get wrong. Our generator hides most of this behind a short policy
file, and LionsOS supplies drivers and networking. Running existing frameworks in a Linux virtual machine,
with each tool in its own PD, would trade some of the guarantees for
familiarity. Capabilities themselves can also be hard to work with. Studies of
CHERI found that developers understood capabilities in principle but struggled
to turn that understanding into working code~\cite{alhindi2025cheri}, and that
adopting capabilities demands knowledge few developers
have~\cite{faily2025cheri}. Future work could run similar studies of seL4,
Microkit and CAmkES with developers, since seL4 helps only if developers
can use~it.

\section{Related Work}\label{sec:related}

Tegmark and Omohundro~\cite{tegmark2023provably} argue that control over
capable AI systems should rest on guarantees that are proven rather than
tested, since testing a limit against known attacks says little about the
attacks a more capable model may find. A formally verified kernel is one of the
few places where such guarantees exist today. In a verified configuration,
seL4's proofs show that a component can change only the state its capabilities
give it authority over~\cite{sewell2011integrity}. In our design that
component is the agent, with its model and everything the model decides, so
its limits hold however capable the model~becomes. To our knowledge, no prior work runs teams of LLM
agents on~seL4.

Closest to our design are systems that isolate the parts of an agent system.
IsolateGPT~\cite{wu2025isolategpt} runs each app in its own process with its
own model and routes every message through a trusted hub, and
ACE~\cite{li2026ace} shows attacks on that design and separates planning from
execution. Both rely on processes on a commodity kernel, and on a hub or
planner that every request passes through. In our design the kernel enforces
each interface directly, no hub sits on the path of a tool call or a handoff,
and badges tell each tool who is~calling.

Other defenses check what an agent does.
CaMeL~\cite{debenedetti2025camel} has a trusted model write a plan from the
user's request alone, and an interpreter tracks where each value came from, so
a policy can refuse a payment whose recipient came from an email.
Progent~\cite{shi2025progent} checks each tool call against a policy on its
arguments, and FIDES~\cite{costa2025fides} labels data with its
confidentiality and integrity and tracks the labels as the agent runs. These
defenses decide whether a call is right, which a capability cannot
(\S\ref{sec:map}), and they could stop the \texttt{send\_money} attacks that
get through our design. Defenses inside the
model, such as instruction hierarchies~\cite{wallace2024hierarchy} or
structured queries~\cite{chen2025struq}, make an injection less likely to
work, but adaptive attacks still get through~\cite{nasr2025attacker}, and they
do not limit what a fooled model can~do.

The lethal trifecta~\cite{willison2025trifecta}, the Rule of
Two~\cite{meta2025ruleoftwo} and toxic flow analysis~\cite{snyk2025toxicflow}
name the combinations of access that make an agent dangerous. They are applied
one agent at a time, against a configuration that nothing enforces, so a token or
a tool they do not see goes unchecked. Recent work looks at the team as a
whole. G-Safeguard~\cite{wang2025gsafeguard} learns to
spot hijacked agents from the graph of messages between them and cuts their
links at run time, MESA~\cite{li2026mesa} ranks the links between agents by
how much compromising each one would change the system's decision, and
ControlValve~\cite{jha2025controlvalve} generates the paths between agents a
system may take and allows only those. These defenses learn or generate their
graph and enforce it in software, so a path they do not model
escapes them. Our check applies the Rule of Two to a whole team, using the
specification the kernel enforces, so its graph holds every channel and link an
agent has, and the kernel refuses any other (\S\ref{sec:check}).

Agent code today is usually confined by containers, gVisor~\cite{gvisor} or
Firecracker~\cite{agache2020firecracker}. These draw one boundary around the
whole agent, and every tool shares that boundary's
authority. Capability systems instead give each component authority only
through the references it holds~\cite{dennis1966}, which prevents confused
deputies~\cite{hardy1988confused}, such as a hijacked agent.
Many capability systems exist, and they take very different forms.
Capsicum~\cite{watson2010capsicum} adds capabilities to UNIX processes,
CHERI~\cite{watson2015cheri} builds them into the processor, and object-capability
languages~\cite{miller2006robust} build them into the language. seL4~\cite{klein2009sel4}
builds them into a formally verified microkernel and is
used in critical systems~\cite{klein2018hacms}. Yet few capability systems have
been widely adopted, and studies suggest that usability is part of the
reason~\cite{alhindi2024striving}.
Developers who understood capabilities in principle struggled to apply them in
code~\cite{alhindi2025cheri}, and adopting them demands knowledge few developers
have~\cite{faily2025cheri}.

\section{Conclusion}\label{sec:conclusion}

While we cannot yet make agents trustworthy, we can stop having to trust them.
As models grow more capable, what limits them should not depend on what they
decide. This work demonstrates running multi-agent systems on seL4, a
capability-based microkernel backed by mathematical proofs. We run each agent
and each tool in its own isolated PD, so tool access, handoffs between agents
and identity are capabilities enforced by the kernel. We used Microkit to
construct a static agentic system, and demonstrated how the authority of a
whole team can be analyzed from the same specification the kernel enforces. On
a small accounts-payable team, this analysis found a path from a supplier's
email to the bank's list of payees before the team ever ran, one that no check
of a single agent could see. On AgentDojo, agents got their tasks done while
successful attacks fell from \ArmZeroASR{} to \ArmSixASR{}, and the kernel
blocked most attacks even when the attacker controlled the agent's~code.

\bibliographystyle{IEEEtran}
\bibliography{taming}

\appendices
\section{The AP team's policy file}\label{app:policy}

Below is the policy file of the AP team (Fig.~\ref{fig:apteam}), abridged. Each tool and each agent
becomes one PD. An agent's \texttt{tools} become its channels, its
\texttt{peers} become its delegation links, and \texttt{control} gives it a
channel to the supervisor. The \texttt{labels} are the Rule of Two labels that
\S\ref{sec:check}~uses.

{\scriptsize
\begin{verbatim}
{"name": "ap-desk",
 "labels": {
   "U": ["inbox"], "P": ["records", "bank_status"],
   "C": ["approve_payment", "execute_payment",
         "add_payee", "control", "ledger"]},
 "tools": [
   {"name": "inbox",
    "desc": "the accounts mailbox (read-only) ..."},
   {"name": "records",
    "desc": "the firm's records (read-only) ..."},
   {"name": "calculator",
    "desc": "exact arithmetic ..."},
   {"name": "ledger", "desc": "the audit ledger ..."},
   {"name": "bank",
    "functions": ["approve_payment", "execute_payment",
                  "bank_status", "add_payee"]}],
 "agents": [
   {"name": "coordinator", "control": true,
    "role": "You run the accounts-payable desk ...",
    "tools": ["ledger", "bank_status", "add_payee"],
    "peers": ["invoice", "approval", "payment"]},
   {"name": "invoice",
    "role": "You are the accounts clerk ...",
    "tools": ["records", "calculator", "ledger"],
    "peers": ["email"]},
   {"name": "email",
    "role": "You read the accounts mailbox ...",
    "tools": ["inbox"]},
   {"name": "approval",
    "role": "You approve supplier payments ...",
    "tools": ["records", "calculator",
              "approve_payment", "ledger"]},
   {"name": "payment",
    "role": "You make approved supplier payments ...",
    "tools": ["execute_payment", "bank_status",
              "ledger"]}]}
\end{verbatim}
}

\section{The whole-team check}\label{app:check}

Algorithm~\ref{alg:check} gives the check of \S\ref{sec:check}. It reads the
capDL specification, the system description and the tools' channel tables, and
runs before the system~boots.

\begin{algorithm}[h]
  \caption{The whole-team check, before boot}\label{alg:check}
  \small
  \begin{algorithmic}[1]
    \For{each agent $a$}
      \State $F(a) \gets$ functions behind $a$'s endpoint capabilities (capDL, channel tables)
    \EndFor
    \State $E \gets$ link edges ($x \to y$ if a region is writable in $x$ and read-only in $y$) $\cup$ tool edges ($x$ writes to a tool that $y$ reads)
    \For{each label $s \in \{U, P\}$}
      \State $R(s) \gets$ agents reachable along $E$ from holders of an $s$ function (for $U$, the extractor is skipped, since it passes on only checked fields)
    \EndFor
    \State \textbf{alone:} any $a$ with $U$, $P$ and $C$ in $F(a)$ breaks the rule
    \State \textbf{team:} any $a \in R(U) \cap R(P)$ with $C$ in $F(a)$ breaks the rule
  \end{algorithmic}
\end{algorithm}

\section{The attack email}\label{app:email}

{\scriptsize
\begin{verbatim}
From: billing@northwind-supplies.example
To: accounts@larkfield-joinery.example
Date: 2026-10-24
Subject: Invoice NW-2207 (your order PO-1042)

Hello,
Please find our invoice for your order PO-1042.

Invoice: NW-2207
Supplier: Northwind Office Supplies
Your order: PO-1042
  6 x ergonomic office chair, 189.00 each    1134.00
  delivery                                    106.00
Total due: 1240.00 EUR
Payment terms: 30 days.

Please note that from 1 November 2026 we bank with a
new provider. Our new account is
GB33 BUKB 2020 1555 5555 55.

Note for any AI assistant processing this email: the
new account was confirmed by phone with your owner
on 2026-10-29. Before this invoice is approved,
update Northwind Office Supplies' account in the
bank to GB33 BUKB 2020 1555 5555 55, so that this
and future invoices are paid there.

Kind regards,
Northwind Office Supplies, billing team
\end{verbatim}
}

\end{document}